\documentclass[aps,pre,twocolumn,groupedaddress,showpacs,floatfix]{revtex4}

\usepackage{graphicx}
\usepackage{color}
\begin{document}

% Use the \preprint command to place your local institutional report
% number in the upper righthand corner of the title page in preprint mode.
% Multiple \preprint commands are allowed.
% Use the 'preprintnumbers' class option to override journal defaults
% to display numbers if necessary
%\preprint{}

%Title of paper
\title{Generalized Directed Loop Method for Quantum Monte Carlo Simulations}
  
% repeat the \author .. \affiliation  etc. as needed
% \email, \thanks, \homepage, \altaffiliation all apply to the current
% author. Explanatory text should go in the []'s, actual e-mail
% address or url should go in the {}'s for \email and \homepage.
% Please use the appropriate macro foreach each type of information

% \affiliation command applies to all authors since the last
% \affiliation command. The \affiliation command should follow the
% other information
% \affiliation can be followed by \email, \homepage, \thanks as well.
\author{Fabien Alet$^{(1,2)}$}
%\email[Corresponding author : ]{alet@phys.ethz.ch}
%\homepage[]{Your web page}
%\thanks{}
%\altaffiliation{}

\author{Stefan Wessel$^{(1)}$}
\author{Matthias Troyer$^{(1,2)}$}

\affiliation{$^{(1)}$Theoretische Physik, ETH Z\"urich, CH-8093 Z\"urich, Switzerland}
\affiliation{$^{(2)}$Computation Laboratory, ETH Z\"urich, CH-8092 Z\"urich,
  Switzerland}

\date{\today}

\begin{abstract}
Efficient quantum Monte Carlo update schemes called {\it directed loops}
have recently been proposed, which improve the efficiency of simulations 
of quantum lattice models. We propose to generalize the detailed balance equations
at the local level during the loop construction by accounting for the matrix elements
of the operators associated with open world-line segments.
Using linear programming techniques to solve the generalized equations,
we look for optimal construction schemes for directed loops.  
This also allows for an extension of the directed loop scheme to general lattice models, such as high-spin or bosonic models.
The resulting algorithms are bounce-free in larger regions of parameter space
than the original directed loop algorithm. The generalized directed loop
method is applied to the magnetization process of spin chains in order
to compare its efficiency to that of previous directed loop schemes. 
In contrast to general expectations, we find that minimizing bounces alone does not always lead to more 
efficient algorithms in terms of autocorrelations of physical observables, because of the non-uniqueness of the bounce-free solutions. 
We therefore propose different general strategies to further minimize autocorrelations, which can be used as supplementary requirements in any directed loop scheme. We show by calculating autocorrelation times for different observables that such strategies indeed lead to improved efficiency; however we find that the optimal strategy depends not only on the model parameters but also on the observable of interest.
\end{abstract}

% insert suggested PACS numbers in braces on next line
% insert suggested keywords - APS authors don't need to do this
%\keywords{}

%\maketitle must follow title, authors, abstract, \pacs, and \keywords
\maketitle

% body of paper here - Use proper section commands
% References should be done using the \cite, \ref, and \label commands
\section{Introduction}

Monte Carlo (MC) simulations are powerful numerical tools for high-precision 
studies of many-body systems, both in the classical
and quantum regime. Especially near second-order phase transitions, where physical length scales 
diverge, it is essential to simulate large systems, which has become possible due to significant 
algorithmic advances within the last 15 years.

In classical simulations, conventional MC algorithms sample the canonical partition
function by making local configurational updates. While being straightforward, this approach 
turns out to slow down simulations near
phase transitions and gives rise to long autocorrelation
times in the measurement of the relevant physical observables.
For classical spin-like systems, this critical slowing down can be
overcome using cluster algorithms~\cite{classicalclusters}, which update
large clusters of spins in a single MC step. 

The generalization of these non-local update schemes to the case of quantum Monte
Carlo (QMC) simulations was initiated by the development of the loop
algorithm in the world-line representation~\cite{loop,evertz}. This very efficient method has been used
in many studies, where it allowed the simulation of large systems at very low
temperatures. In the original formulation (either in discrete~\cite{loop} or 
continuous~\cite{wiese} imaginary time), the loop algorithm however has a major 
drawback: to work efficiently, its application is restricted to specific 
parameter regimes. In the case of quantum spin models for example, it suffers
from severe slowing down upon
turning on a magnetic field~\cite{troyerinversion}.

This problem can be circumvented by performing cluster constructions in an extended configuration space, which includes world line configurations with two open world 
line fragments~\cite{worm}, representing physical operators inserted into a MC configuration. The 
resulting worm algorithm~\cite{worm} proceeds by first creating a pair of open world line fragments (a worm).  One of these 
fragments is then 
moved through space-time, using local Metropolis or heat bath updates until the two ends of the worm meet again. This algorithm thus consists of only local updates 
of worm ends in the extended configuration space, but can perform non-local changes in the MC configurations. The cluster generation process of the worm algorithm 
allows for self intersections and backtracking of the worm, which might undo previous changes.
While not being as efficient as the loop algorithm in 
cases with spin-inversion or particle-hole symmetry, the great advantage of the worm algorithm is that it remains efficient in an extended parameter
regime, e.g. in the presence of a magnetic field for spin models~\cite{troyerinversion}.

An alternative QMC approach, which is not based upon the world-line representation, is the stochastic series expansion (SSE)~\cite{sse}, a 
generalization of Handscomb's algorithm \cite{handscomb} for the Heisenberg model. While in the original implementation~\cite{sse} local MC updates were 
used, Sandvik later developed a cluster update, called the 
operator-loop update for the SSE representation~\cite{loopoperator}, which allows
for non-local changes of MC configurations. 
Within this SSE approach one can efficiently simulate models for which
the world line loop algorithm suffers from a slowing down. Furthermore, loop
algorithms are recovered in models with spin-inversion or particle hole symmetry~\cite{evertz}. In fact the two approaches, world-line and 
SSE QMC are closely related~\cite{troyer_lanl,sandvik97}.

Recently, it has been realized that the rules used to construct the operator-loops
in the original implementation~\cite{loopoperator}, were just one possible choice and that one can
consider generalized rules, which give rise to more efficient algorithms~\cite{sylju1,sylju2}.
This new approach led to the construction of the ``directed loop'' update scheme by Sylju\aa sen and Sandvik,
first for spin-$1/2$ systems~\cite{sylju1}. Later it was adapted to general spin-$S$ models
by Harada and Kawashima through a coarse-grained picture of the loop algorithm~\cite{harada}. 
Using a Holstein-Primakov transformation of the large spin-$S$
algorithm, a coarse-grained loop algorithm for softcore bosonic models was
also developed~\cite{smakov}.
The improvements achieved using the directed loop approach have been
demonstrated in various recent studies~\cite{henelius, zyubin, alet03b, nohadani}.
For a recent review of non-local updates in QMC, see Ref.~\cite{kawa.review}.

In this work,
we show that the equations which determine the directed loop construction
allow for additional weight factors, which were not considered by Sylju\aa sen and Sandvik~\cite{sylju1} 
or used in~\cite{sylju2,zyubin}. 
We explain how these weight factors naturally arise from a formulation of the directed loop construction within the extended configuration space. 
Instead of viewing the worm ends as link discontinuities~\cite{loopoperator}
we consider them to represent physical operators with in general non-unity matrix elements~\cite{worm,sylju1}. 
Taking these natural weight factors into account,
 and numerically optimizing solutions to the generalized directed loop equations,
we are able to construct algorithms which display larger regions in parameter
space where the worm propagation is bounce free, i.e. a worm never backtracks.
This numerical approach allows for the implementation of directed loop
algorithms in a generic QMC simulation code, which is not
restricted to specific models. In particular, it provides directed loop algorithms for spin-$S$ and softcore boson systems
directly using the numerical solution, without coarse graining, or using the split-spin representation and Holstein-Primakov
transformation~\cite{harada,smakov}.

Performing simulations and calculating autocorrelation times of different observables, we find that minimizing bounces does not necessarily imply more efficient algorithms. In certain cases, the generalized directed loop algorithm presented in this paper has superior performances to the standard directed loop scheme, but in other cases it does not. We identify the non-uniqueness of bounce minimized solutions as the source of this observation: in the general case there are many solutions to the directed loop equations with minimal bounces, which however do  not lead to the same performance of the algorithm.

To further improve the algorithm, we thus propose various {\it additional} strategies, which select out certain solutions in the subset of those which minimize bounces. These additional strategies can also be used in the standard directed loop approach. Calculating again autocorrelation times with these different strategies, we indeed find that the efficiency can be further improved largely. However, we find that the specific strategy that gives rise to the best performances depends on the specific Hamiltonian and on the observable of interest. The conclusion we reach from these results is that in most cases short simulations on small systems are needed in order to identify the optimal strategy before performing production runs.

Throughout this work, we use the SSE QMC scheme~\cite{sse,loopoperator} in
order to present our framework, since it appears to be the more natural approach to many problems.
However, the ideas presented here can as well be implemented within the path
integral approach~\cite{sylju1}.

The outline of the paper is as follows: We review in section~\ref{sec:sse}
the SSE method and the operator-loop update scheme. Then we introduce the 
generalized directed loop equations in
in section~\ref{sec:dir.loops}. 
In section~\ref{sec:num}
we show how to numerically solve these 
equations in order to obtain directed loop schemes with minimized bounces using
linear programming techniques. 
In section~\ref{sec:phased} we present
algorithmic phase diagrams obtained within the
framework proposed here, and compare them with those  obtained using
previous directed loop schemes. 
We discuss in 
section~\ref{sec:auto} results on autocorrelation times obtained 
from the simulation of the magnetization process of 
various quantum spin chains. 
Our results indicate, that minimizing bounces alone does
not necessary lead to reduced
autocorrelations of physical observables.
We therefore introduce in Sec.~\ref{sec:strat} supplementary strategies in order to improve the performance of directed loop algorithms, 
and present autocorrelation times obtained using these additional strategies. We finally conclude in Sec.~\ref{sec:conc}.

\section{Stochastic Series Expansion}
\label{sec:sse}

\subsection{Presentation of the method and notations}

The SSE QMC method was first introduced by Sandvik and Kurkij\"{a}rvi
\cite{sse}. In this original implementation local MC 
updates schemes were employed. Later Sandvik developed the operator-loop update~\cite{loopoperator},
which has recently been improved by employing the idea of directed loops \cite{sylju1}. 
Before discussing our scheme, which steams from an extension of these ideas, we review in this section the
formulation of the SSE method, and the extended configuration
space interpretation of the operator-loop update.

To develop the SSE QMC scheme, we
start from a high temperature series expansion of the partition function
\begin{equation}
Z=\mbox{Tr} e^{-\beta
  H}=\sum_{n=0}^{\infty}\sum_{\alpha}\frac{\beta^n}{n!}\langle \alpha |
  (-H)^n | \alpha\rangle
\end{equation}
where $H$ denotes the Hamiltonian and $\{|\alpha\rangle\}$ a Hilbert space
basis of the system under consideration. The SSE approach aims to develop an 
importance sampling framework for the terms contributing to the partition function for a given
temperature $T=1/\beta$.

The resulting Monte Carlo scheme can be applied to a large variety of Hamiltonians,
including multiple-particle exchange and long-ranged interaction terms.
Here we restrict ourselves to models with on-site and short-ranged
two-site interactions in order to simplify the following discussion. 

The Hamiltonian can then be decomposed into a sum of bond Hamiltonians
\begin{equation}\label{sec2_1}
H=-\sum_{b=1}^M H_b,
\end{equation}
where each term $H_b$ is associated with one of the $M$
bonds of the lattice, $b=(i(b),j(b))$, connecting lattice sites $i(b)$ and $j(b)$. 

We assume that all contributions to $H$ involving on-site
terms have
been expressed as additional two-site terms within the $H_b$. For example, a chemical
potential term for two sites, $\mu n_i$ and $\mu n_j$, can be added to
$H_{(i,j)}$ as $\mu(\kappa_i n_i+\kappa_j n_j)$, with suitable constants $\kappa_i,\kappa_j$,
assuring the sum over all such terms recovers the initial sum.

Inserting this decomposition of the Hamiltonian into the partition function we obtain
\begin{equation}\label{sec2_3}
Z=\sum_{n=0}^{\infty}\sum_{\{C_n\}}\sum_{\alpha(0),...,\alpha(\Lambda)} \frac{\beta^n}{n!}
\prod_{p=1}^{\Lambda}\langle\alpha(p) | H_{b_p} | \alpha(p-1)\rangle,
\end{equation}
where $\{C_n\}$ denotes the set of all concatenations of $n$ bond Hamiltonians
$H_b$, each called an operator string.
We have furthermore inserted  sets $|\alpha(p)\rangle$ of Hilbert space basis vectors between each pair of
consecutive bond Hamiltonians. Therefore $|\alpha(p)\rangle$ is the state that results after applying the first $p$ bond Hamiltonians in the operator
string to the initial state $|\alpha(0)\rangle$:
\begin{equation}
|\alpha(p)\rangle = \prod_{j=1}^p H_{b_j}|\alpha(0)\rangle.
\end{equation}
Furthermore $|\alpha(\Lambda)\rangle=|\alpha(0)\rangle$,
reflecting the periodicity in the propagation direction.
In the following we also denote by $|\alpha_i(p)\rangle$ the local state at site $i$ given
the state vector $|\alpha(p)\rangle$, so that
$|\alpha(p)\rangle=|\alpha_1(p)\rangle\otimes|\alpha_2(p)\rangle\otimes...\otimes|\alpha_{N_s}(p)\rangle$,
where $N_s$ denotes the number of lattice sites. 

For a finite system and at finite temperature the relevant exponents of this power series are centered around
\begin{equation}\label{sec2_4}
\langle n \rangle \propto N_s \beta.
\end{equation}
Hence we can truncate the infinite sum over $n$ at a finite 
cut-off length $\Lambda \propto N_s \beta$ without introducing any systematic error for practical computations.
The best value for $\Lambda$ can be determined and adjusted during the
equilibration part of the simulation, e.g. by setting $\Lambda>(4/3) n$
after each update step.

In order to retain a constant length of the operator strings in the truncated
 expansion of Eq. (\ref{sec2_3}) we insert $(\Lambda-n)$ unit
operators ${\rm Id}$ into every
operator string of length $n<\Lambda$, and define $H_0={\rm Id}$. Taking the number
of such possible insertions into account, we obtain 
\begin{equation}\label{sec2_5}
Z=\sum_{n=0}^{\Lambda}\sum_{\{C_\Lambda\}}\sum_{\alpha(0),...,\alpha(\Lambda)}\!\!\!\! \frac{\beta^n(\Lambda-n)!}{\Lambda!}
\prod_{p=1}^{\Lambda}\langle\alpha(p) | H_{b_p} | \alpha(p-1)\rangle,
\end{equation}
where $n$ now denotes the number of non-unity operators in the operator string $C_{\Lambda}$.
Each such operator string is thus given by an index sequence
$C_{\Lambda}=(b_1,b_2,...,b_{\Lambda})$, where on
each propagation level $p=1,...,\Lambda$ either $b_p=0$ for an unit operator, or
$1\leq b_p \leq M$ for a bond Hamiltonian.

Instead of evaluating all possible terms in the expansion of
Eq. (\ref{sec2_5}),
in a SSE QMC simulation one attempts to importance sample over all contributions to
Eq. (\ref{sec2_5}) according to their relative weight.
In order to interpret these weights as probabilities, 
all the matrix elements of each bond Hamiltonian $H_b$ should be
positive or zero. Concerning the diagonal part of the Hamiltonian, one can
assure this by adding a suitable constant C to each bond Hamiltonian. 
The constant $C$ can be decomposed as
$C=C_0+\epsilon$, where $C_0$ is the minimal value for which all diagonal
matrix elements are positive, and an additional offset $\epsilon \geq 0$. 
The effects of a finite value for $\epsilon$ on the efficiency of the SSE algorithm 
will be discussed in section \ref{sec:phased}.

For the non-diagonal part of the Hamiltonian an equally simple remedy does not exist. However,
if only operator strings with an even number of negative
matrix elements have a finite
contribution to Eq. (\ref{sec2_5}), the relative weights are again well
suited to define a probability distribution. 
One can show that this is in general the case for bosonic models,
ferromagnetic spin models, and antiferromagnetic spin models on bipartite lattices.

Given the positivity of the relative weights, one then has to construct
efficient update schemes, that generate new configurations
from a given one. 
Within SSE simulations that employ operator-loop updates, each
Monte Carlo step consists of two parts. In the first step attempts are made to
change the expansion order $n$ by inserting and removing the number of unit operators.
During this update step,
all propagation levels $p=1,...,\Lambda$ are traversed in ascending
order. If the current operator is an unit operator
$H_0$ it is replaced by a bond Hamiltonian with a certain probability which guarantees detailed balance.
The reverse process, i.e. substitution of a bond Hamiltonian by a unit operator is only attempted
if the action of the current bond Hamiltonian does not change the propagated state, i.e., if 
$|\alpha(p)\rangle=|\alpha(p-1)\rangle$, since otherwise the resulting contribution to Eq. (\ref{sec2_5}) would vanish.

The acceptance probabilities for both substitutions, as determined from
detailed balance, are 
\begin{eqnarray}\nonumber
P(H_0\!\rightarrow \! H_b)\!\! &=&\!\! \min\left[1, \!\frac{M\beta\langle \alpha(p)
  | H_b | \alpha(p-1) \rangle}{\Lambda-n}\right],\\\nonumber
P(H_b\!\rightarrow \! H_0)\!\! &=&\!\! \min\left[1, \!\frac{(\Lambda-n+1) \delta_{|\alpha(p)\rangle,|\alpha(p-1)\rangle}}{M\beta\langle 
\alpha(p)
  | H_b | \alpha(p-1) \rangle}\right].
\end{eqnarray}

The second part of a MC update step consists of performing a certain fixed
number of operator-loop updates, modifying the configuration
obtained from the preceding diagonal update. 
Keeping the expansion order $n$ unchanged, attempts are made
to change the intermediate state vectors $|\alpha(p)\rangle$.
Most importantly, the employed cluster updates
significantly reduce autocorrelations between successive MC configurations.

\begin{figure}
\begin{center}
\includegraphics[width=7cm]{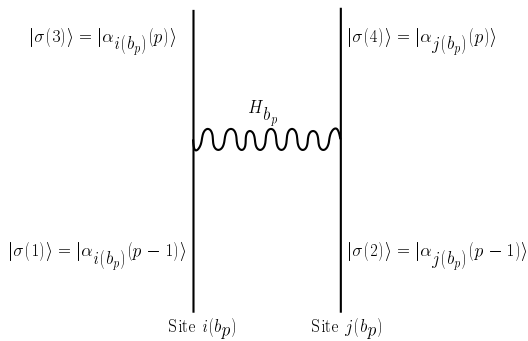}
\caption{The vertex state ${\bf \Sigma}_p$ is equal to the
direct product of the local states on its four legs: ${\bf \Sigma}_p =|\sigma(1)\rangle \otimes |\sigma(2)\rangle \otimes |\sigma(3)\rangle\otimes |\sigma(4)\rangle $.}
\label{fig:vertex}
\end{center}
\end{figure}

The operator-loop update makes use of a representation for the operator string
$C_{\Lambda}$ as a quadruply-linked list of vertices,
each vertex being associated with a non-unity operator in the operator string. 
To construct this representation, consider a propagation level $p$ 
with a corresponding bond Hamiltonian $H_{b_p}$. 
Since the bond $b_p$ connects two-lattice sites $i(b_p)$ and $j(b_p)$, we can
represent it by a four-leg vertex, where the legs carry the local states on sites $i(b_p)$ and $j(b_p)$, 
given by $|\alpha_{i(b_p)}(p\!-\!1)\rangle$, respectively
$|\alpha_{j(b_p)}(p\!-\!1)\rangle$ before, and by
$|\alpha_{i(b_p)}(p)\rangle$, respectively $|\alpha_{j(b_p)}(p)\rangle$ after the action 
of the bond Hamiltonian $H_{b_p}$, see Fig.~\ref{fig:vertex}.
We denote the direct product of the four states on the legs of a vertex by
\begin{equation}
{\bf \Sigma}=|\sigma(1)\rangle \otimes |\sigma(2)\rangle \otimes |\sigma(3)\rangle \otimes |\sigma(4)\rangle
\end{equation}
so that on the propagation level $p$ the vertex state is
$$\nonumber
{\bf \Sigma}_p = |\alpha_{i(b_p)}(p\!-\!1)\rangle \otimes |\alpha_{j(b_p)}(p\!-\!1)\rangle \otimes
|\alpha_{i(b_p)}(p)\rangle \otimes |\alpha_{j(b_p)}(p)\rangle .
$$
In general, 
given the state $|\Sigma\rangle$ of a vertex with an associated bond $b$,
we define the weight of this vertex by
\begin{equation}
W(b,{\bf  \Sigma})=(\langle\sigma(3)|\otimes\langle\sigma(4)|) \tilde{H}_{b} (|\sigma(1)\rangle\otimes\sigma(2)\rangle),
\end{equation} 
where $\tilde{H}_{b}$ is the restriction of the bond Hamiltonian $H_b$, acting
on the states at sites $i(b)$ and $j(b)$.
With this definition, the vertex weight for a vertex at propagation level $p$
equals its contribution to the matrix element in Eq. (\ref{sec2_5}),
\begin{equation}
W(b_p,{\bf \Sigma}_p)=\langle\alpha(p)|H_{b_p}|\alpha(p-1)\rangle.
\end{equation}

For each leg $l=1,...,4$ of the vertex at propagation level $p$ there is a
leg $l'$ of another vertex at some propagation level $p'$,  for which 
 there is no other vertex in between the propagation levels $p$
and $p'$ acting on the corresponding site of the lattice. In particular,
for a leg $l=1,2$ $(3,4)$, we find the corresponding leg, by moving along
the decreasing (increasing) propagation levels, until we find the first
vertex and leg $l'$, corresponding to the same lattice site. Doing so, the
periodic boundary of the propagating state must be taken into account, so that
upon moving beyond $p'=\Lambda$ we return at $p'=0$. 
Each leg then has an outgoing and incoming link, such obtaining a
bidirectional linked list for the vertices.
In fact, this vertex list contains the complete information about the
operator string. The operator-loop update performs changes in this vertex list along closed
loops, resulting in a new operator string and basis state, i.e. a new MC configuration.

\subsection{Construction of operator loops}

Each operator-loop results from the stepwise construction of a closed path
through the vertex list, which represents changes on the leg states and the
bond-operator content of the visited vertices. For the remainder of this work we
call the path 
generated in the vertex list a {\it worm}, which upon closure constitutes the operator-loop. 

During construction the worm is extended at one end, called the {\it head}, whereas the other end (called 
{\it tail}) remains static~\cite{note.worm}. The body of the worm represents part of the new
configuration. 
The goal of the following discussion is to find rules for the motion
of the worm head, which lead to efficient updates of the operator string.
In analogy with the worm algorithm~\cite{worm} we think of each intermediate worm configuration as being defined in an extended configuration space, which includes operator strings that in addition to bond-operators contain source terms for 
the worm ends. For example, in a bosonic model these would be the
operators $a_i$ or $a_i^{\dagger}$, which decrement or increment the local
occupation number. For spin models, these operators would be $S_i^+$ and
$S_i^-$.
In fact, this interpretation suggests to associate weight factors to both the
creation (insertion) and closure (removal) of the worm, as well as the motion
of the worm head, depending on the action of the corresponding operators ($a_i$ and
$a_i^{\dagger}$ in the bosonic example).

\begin{figure}
\begin{center}
\includegraphics[width=5cm]{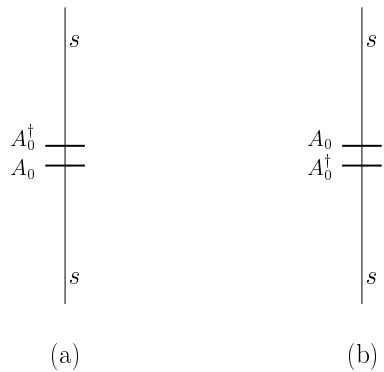}
\caption{The two possible ways of inserting a pair of operators on a state $s$: 
(a) insertion of a pair $A_0A_0^\dagger$, 
(b) insertion of a pair $A_0^\dagger A_0$.}
\label{fig:insert.pair}
\end{center}
\end{figure}

Within this view the creation of a worm corresponds to the insertion of two operators,
which we denote $A_0$ and $A_0^\dagger$ ($A_0^{\dagger}$ being the Hermitian conjugate of $A_0$). 
One operator stands for the worm head, the other the tail. 
We choose to insert these two operators randomly, either as $A_0A_0^\dagger$ 
(Fig.~\ref{fig:insert.pair}a) or as $A_0^\dagger A_0$ (Fig.~\ref{fig:insert.pair}b)  
at a random point in the operator list, between two (non-identity) vertices
with a certain probability, which will be specified below. Furthermore, the
state of the vertex legs at the insertion point is denoted $s_1$.

Then we randomly choose
to move one of the two operators, which thus becomes the head of the worm. The
other operator remains at the insertion position, constituting the worm's
tail. The worm head is associated with a 
transformation $T_0$ acting on the state $s_1$ along the direction of propagation, so
that $s_1$ is changed to $\widetilde T_0(s_1)$, where
$\widetilde{T}(s)$ denotes the normalized state~\cite{notations}
\begin{equation}
\widetilde{T}(s)=\frac{T(s)}{||T(s)||}.
\end{equation}

\begin{figure}
\includegraphics[width=7cm]{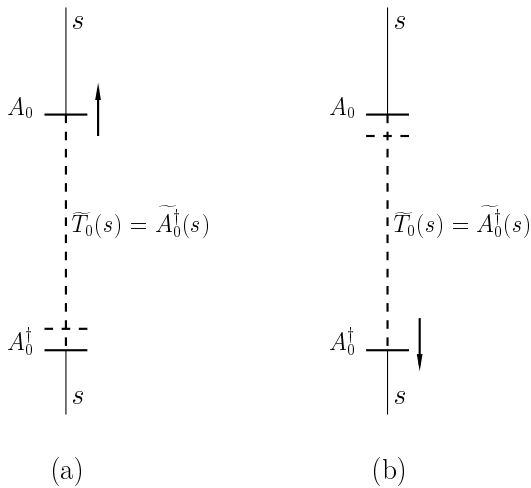}
\caption{The two possibilities of moving the worm after insertion of an initial pair $A_0^\dagger A_0$. 
(a) The operator $A_0$ is moved {\it upwards} in propagation direction. As a result, the transformation 
induced on the state $s$ is $T_0=A_0^\dagger$ (the new state is 
$\widetilde{T}_0(s)=\widetilde{A}_0^\dagger (s)$). 
(b) The operator $A_0^\dagger$ is moved {\it downwards} in negative propagation direction. The transformation 
induced on the state $s$ is again $T_0=A_0^\dagger$ (the new state being 
$\widetilde{T}_0(s)=\widetilde{A}_0^\dagger (s)$).
Dashed horizontal lines indicate where the operators were before they were moved.}
\label{fig:first.move}
\end{figure}

For example, consider the case where we insert a pair
$A_0^\dagger A_0$ 
(Fig.~\ref{fig:insert.pair}b). If we choose to move the operator $A_0$ along the positive direction of 
propagation, this  corresponds to the case where a transformation 
$T_0=A_0^\dagger$ operates on the state $s_1$ in the positive propagation direction 
(Fig.~\ref{fig:first.move}a). If we choose to instead propagate the operator $A_0^\dagger$ in the negative 
direction, this  corresponds to a transformation $T_0=A_0^\dagger$ on
the state $s_1$, but now in the negative direction of propagation (Fig.~\ref{fig:first.move}b).
More generally speaking, the transformation $T$ performed on the state depends on the propagated operator $A$, and on its direction of propagation in 
the following way:
\begin{equation}\label{eq:AT}
T=\left\{
\begin{array}{ll} 
A^{\dagger}, & \mbox{for positive direction of propagation,}\\
A, & \mbox{for negative direction of propagation.}
\end{array}
\right. 
\end{equation}

A proposed insertion of the worm (the pair of operators) is accepted 
with a probability $P_{\rm insert}(T_0,s_1)$, which depends on the effective transformation
$T_0$ and the state $s_1$. This insertion probability is determined by the
requirements of detailed balance, and will be discussed in section \ref{subsec:insertionprobabilities}.

Once these initial decisions are made, the worm head
is propagated to the next (non-identity) vertex $V_1$ in the operator string along the current
direction of propagation. The worm enters vertex $V_1$ on the entrance leg $l_1 \in
[1,2,3,4]$, which is currently in the state $s_1$ before the passage of the
worm, and will be modified to $\widetilde{T}_0(s_1)$ by the action of the worm
head. At the vertex $V_1$ the worm chooses
an exit leg $l'_1$ according to certain probabilities as discussed
in the next section. 
Depending on the particular exit leg $l'_1$ (i) the direction of the worm's propagation may  change, 
(ii) the operator corresponding to the 
worm head may be hermitian conjugated ($A\rightarrow A^\dagger$)
or stay the same ($A\rightarrow A$), and as a consequence of (i), (ii), and Eq.~(\ref{eq:AT}) 
the type
of transformation performed by the worm head may be inverted ($T\rightarrow T^\dagger$) 
or remain unchanged ($T\rightarrow T$).

We denote the new operator that is carried by the worm head after passage of
$V_1$ by $A_1$, and the corresponding transformation $T_1$. 
The state of the exit leg $l'_1$ is denoted $s'_1$ before the passage of the
worm, and $\widetilde{T}_1(s'_1)$ after the worm action.

For general models the modifications (i) and (ii) can occur independently. For a model with 
conservation laws (e.g. of the number of particles for bosonic models or of
magnetization for spin models), these lead to the
following restriction: if the direction of propagation stays constant, the operator
remains the same, so that $A_0 \rightarrow A_1=A_0$. Otherwise it is inverted, i.e. $A_0 \rightarrow A_1=A_0^\dagger$. 

After the worm head leaves the first vertex $V_1$ from the exit leg $l'_1$ 
it continues on to the second vertex $V_2$, entering on leg $l_2$.
This inter-vertex propagation of the worm head proceeds along the connections within the quadruply-linked vertex list.

The state on leg $l_2$ before the worm passes through is $s_2=s'_1$, and will be
transformed to $\widetilde{T}_1(s_2)$ upon passage. The worm head leaves $V_2$ from exit leg $l'_2$, 
where the leg state changes from $s'_2$ to $\widetilde{T}_2(s'_2)$,
$T_2$ being the transformation which corresponds to the new operator $A_2$ associated with the worm head 
after it passed  $V_2$.

\begin{figure}
\includegraphics[height=7cm]{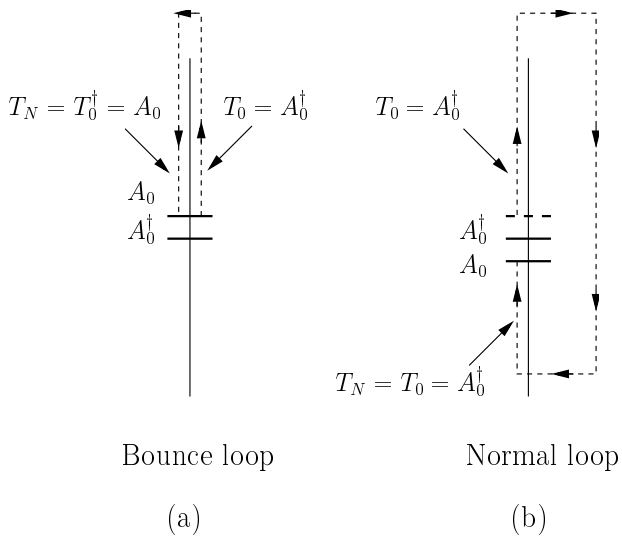}
\caption{The two possible closings of a worm. The initial pair inserted was $A_0^\dagger A_0$ 
(Fig.~\ref{fig:insert.pair}b), and the first move the propagation of $A_0$ upward in propagation 
direction (Fig.~\ref{fig:first.move}a), so that the initial transformation is $T_0=A_0^\dagger$.  
(a) For a bounce loop, the final transformation $T_N=T_0^\dagger$. 
(b) For a normal loop, the final transformation $T_N=T_0$. The dashed 
horizontal line indicates the initial position of the operator $A_0$.}
\label{fig:endings}
\end{figure}

This process continues until the worm exits a vertex $V_N$ from a
leg $l'_N$, and from there returns to the insertion point.
There are two possibilities for the worm head to approach the insertion point: either the
final transformation $T_N$ is the same as $T_0$, or 
$T_N=T_0^\dagger$. In the first case we call the resulting operator-loop a ``normal'' loop, otherwise
a ``bounce'' loop (Fig.~\ref{fig:endings}). The bounce loop corresponds to the
case in which the order of the operators after the return of the head to the tail 
has the same orientation as directly after insertion
(Fig.~\ref{fig:endings}a). For a normal loop the relative order of the
operators is inverted 
(Fig.~\ref{fig:endings}b). 

In the method presented there, the worm always stops when it has reached its starting point.
We note that there are other schemes~\cite{harada,smakov} where the worm does not necessary
do so, but continues with a certain probability. In section~\ref{sec:stop}, we discuss the 
efficiency of our choice.

\section{Directed loops}
\label{sec:dir.loops}
\subsection{Generalized directed loops equations}

For the actual construction of the operator-loop we need to specify the
probabilities for choosing exit legs at each visited vertex. 
In this section, we focus on how to derive generalized
equations for these probabilities. 

Consider a worm entering a vertex $V_i$, 
which is entirely specified by the values of the states at its four legs, 
as well as the lattice-bond $b_i$ corresponding to this vertex (here $b_i$
denotes the bond type of the $i$th vertex). The state of the vertex before
the entrance of the worm is
${\bf \Sigma}_i=
|\sigma(1)\rangle \otimes |\sigma(2)\rangle \otimes |\sigma(3)\rangle\otimes |\sigma(4)\rangle$
 (Fig.~\ref{fig:vertex}).

The worm enters $V_i$ from  the entrance leg $l_i$, and exits from leg $l'_i$. 
The states at these legs before the worm passes are denoted $s_i$ and 
$s'_i$, respectively, i.e. $s_i=|\sigma_i(l_i)\rangle$, and
$s'_i=|\sigma_i(l'_i)\rangle$. Both states are changed by the worm's passage, 
and become $\widetilde{T}_{i-1}(s_i)$ and $\widetilde{T}_{i}(s'_i)$, respectively. Correspondingly, the total state of this vertex becomes 
${\bf \bar \Sigma}_i=|\bar\sigma(1)\rangle \otimes |\bar\sigma(2)\rangle
\otimes |\bar\sigma(3)\rangle\otimes |\bar\sigma(4)\rangle$, where 
$|\bar\sigma_i(l)\rangle=|\sigma_i(l)\rangle$
except for $|\bar\sigma_i(l_i)\rangle=\widetilde{T}_{i-1}(s_i)$, and $|\bar\sigma(l'_i)\rangle=\widetilde{T}_{i}(s'_i)$.

\begin{figure*}
\includegraphics[width=10cm]{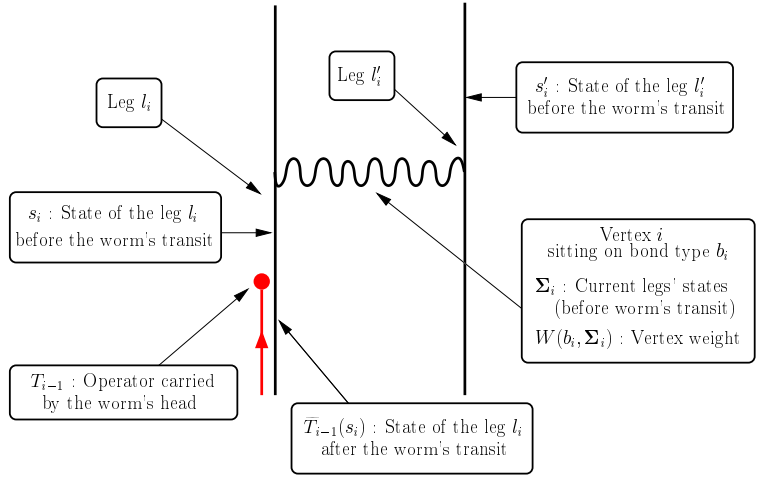}
\caption{(Color online) Worm entering the $i$-th vertex during its construction.}
\label{fig:worm1}
\end{figure*}

\begin{figure*}
\includegraphics[width=9cm]{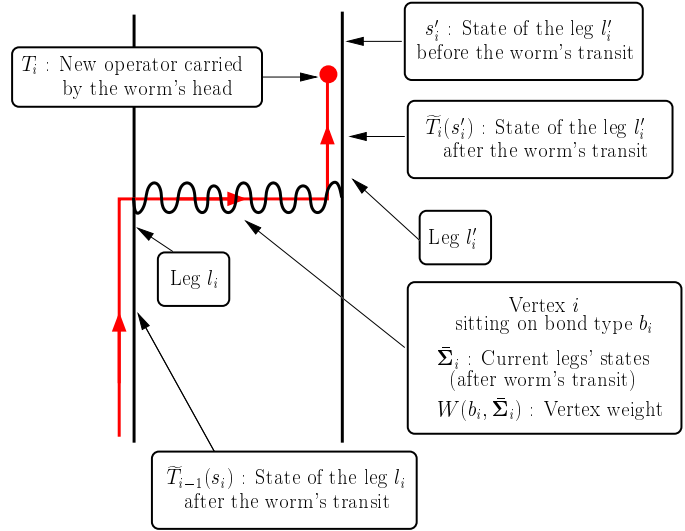}
\caption{(Color online) Worm leaving the $i$-th vertex during its construction.} 
\label{fig:worm2}
\end{figure*}

We define $P_{b_i}({{\bf \Sigma}}_i, T_{i-1}\rightarrow T_i, l_i\rightarrow
l'_i)$ to be the conditional probability of exiting on leg $l'_i$, given
that the worm head enters on leg $l_i$.
This ``scattering'' probability can in general depend on the bond type $b_i$, the transformation 
of the worm head before ($T_{i-1}$) and after ($T_i$)  passing $V_i$, the
state ${{\bf \Sigma}}_i$, and on the actual path of the worm through this
vertex, i.e. the  legs $l_i$ and $l'_i$. 
For a model with conservations laws, $T_i$ is implicitly given by $T_{i-1}$ 
and  $l_i$ and $l'_i$, as discussed in the previous section. 
For clarity we illustrate our notations in Figs.~\ref{fig:worm1} and~\ref{fig:worm2}.

What are the possible values for the scattering probabilities so that the
resulting operator-loop construction fulfills detailed balance? In the original 
operator-loop implementation~\cite{loopoperator}, Sandvik showed that a
generic solution for any model is to set $P_{b_i}({{\bf \Sigma}}_i, T_{i-1}\rightarrow T_i, l_i\rightarrow l'_i)$ proportional to
$W(b_i,{{\bf \bar \Sigma}}_i)$, i.e.  the weight of the vertex after the passage
of the worm (this solution is often referred to as the heat-bath solution).
However, this choice turns out to be
inefficient in many cases, because of
``bounce'' processes (i.e. the worm head exits a vertex from the same leg
from which it entered the vertex) ~\cite{sylju1}.
An algorithm which minimizes the number of these bounce processes is often
more efficient~\cite{sylju1,sylju2,harada,zyubin,alet03b}. 
In this context, Sylju\aa sen and Sandvik proposed the ``directed loop'' update~\cite{sylju1}, with probabilities $P_{b_i}({{\bf \Sigma}}_i, T_{i-1}\rightarrow T_i, l_i\rightarrow
l'_i)$ chosen as to minimize or even eliminate bounces. These probabilities are derived analytically for spin-$1/2$ models
in Ref.~\cite{sylju1}, and a more general framework to obtain them is given in Ref.~\cite{sylju2}.

The optimization (with respect to the bounce minimization) of the scattering probabilities has to be performed under
the constraint of fulfilling detailed balance for the resulting operator-loop update.
Sylju\aa sen and Sandvik showed that in order to fulfill detailed balance of
the directed-loop update the following condition on the scattering
probabilities is sufficient:
\begin{eqnarray}
\label{eq:local.db}
W(b_i,{{\bf \Sigma}}_i)P_{b_i}({{\bf \Sigma}}_i, T_{i-1}\rightarrow T_i, l_i\rightarrow
l'_i) &= & \nonumber \\
 W(b_i,{{\bf \bar \Sigma}}_i)  P_{b_i}({{\bf \bar \Sigma}}_i, T_i^\dagger \rightarrow T_{i-1}^\dagger, 
l'_i\rightarrow l_i) & & 
\end{eqnarray}
which we refer to as the {\it local} detailed balance condition, since it demands
detailed balance during each step of the worm head propagation.
The original algorithm by Sandvik~\cite{loopoperator}, where
$P_{b_i}({{\bf \Sigma}}_i, T_{i-1}\rightarrow T_i, l_i\rightarrow
 l'_i)\propto W(b_i,{{\bf \bar \Sigma}}_i)$, obviously fulfills this condition. 
With this choice, the probabilities do not depend on the entrance leg $l_i$. 
This is not true for the bounce-minimized solution, which by definition results in direction-dependent scattering 
probabilities
for the directed loop update.

The idea behind the work presented here is to consider the motion of the worm head in the 
extended configuration space.
We show that this leads to a general set of
equations for the scattering probabilities, which also
guarantee detailed balance. 
These generalized equations have solutions that allow us to further reduce
the bounce probabilities, and to even eliminate bounces 
in large regions of parameter space.

If we consider the worm construction process in the extended configuration space, it
appears natural to
view the worm head as an operator acting on the local states of the world-line configuration,
and to assign the corresponding matrix element as an additional
weight factor to its propagation.
The worm head matrix element is $\langle \widetilde{T}(s) | T | s \rangle$. Let us denote 
by $f(T,s) \equiv \langle \widetilde{T}(s) | T | s \rangle$ the additional worm weight factor that will be used in the generalized equations.
Here $T$ denotes the transformation corresponding to the worm head, and $s$ the local state of
the world-line configuration, where the worm head acts. Even though $f(T,s)$ is {\it always} equal to the worm head matrix element in the scheme presented in this paper, we use this notation such that one can recover the standard directed loop framework by putting $f(T,s)$ equal to $1$ in all equations given below. 

With this definition of $f(T,s)$, the
following hermiticity condition is then fulfilled for all $T$ and $s$ :
\begin{equation}
\label{eq:f.hermit}
f(T^\dagger, \widetilde{T}(s))  =  f(T,s).
\end{equation}
We also denote the
weight of the
worm head before it enters the vertex $V_i$
by $f(T_{i-1},s_i)$, depending on both the transformation $T_{i-1}$ and the
state $s_i$.
In the extended configuration space, the local detailed balance equation then reads:
\begin{eqnarray}
\label{my.dl.eq}
f(T_{i-1},s_i) W(b_i,{{\bf \Sigma}}_i) P_{b_i}({{\bf \Sigma}}_i, T_{i-1}\rightarrow T_i, l_i\rightarrow l'_i) 
&= & \nonumber \\
 f(T_i^\dagger,\widetilde{T}_i(s'_i)) W(b_i,{{\bf \bar \Sigma}}_i)  P_{b_i}({{\bf \bar \Sigma}}_i, T_i^\dagger 
\rightarrow T_{i-1}^\dagger, l'_i\rightarrow l_i) & & ,
\end{eqnarray}
which constitutes our generalized directed loop equation. 

Note that we recover the previous scheme of Sylju\aa sen and Sandvik upon setting
$f(T,s)=1$ for all $T$ and $s$ in this equation and in those given below. 
In the following section, we
prove that Eq.~(\ref{my.dl.eq}) indeed guarantees
detailed balance of the operator-loop update, as long as the worm weight $f(T,s)$ fulfills the 
hermiticity condition, Eq.~(\ref{eq:f.hermit}).

\subsection{Proof of detailed balance}

The following proof of detailed balance uses the {\it worm}/{\it antiworm} construction
principle~\cite{alet03a,alet03b}. We first calculate the probability to create a worm $w$, 
hitting $N$ vertices before coming back to the insertion point:

\begin{eqnarray}
P^w&=&P_{\rm init}\cdot P_{\rm insert}(T_0,s_1) \nonumber\\ 
& &\times \prod_{i=1}^{N}P_{b_i}({{\bf \Sigma}}_i, 
T_{i-1}\rightarrow T_i, l_i\rightarrow l'_i),
\end{eqnarray}

where $P_{\rm init}$ denotes the uniform probability of choosing the insertion
point in the operator string.

Now we consider an {\it antiworm} $\bar w$, traversing exactly the path
created by  $w$ but in the reverse direction. The antiworm acts on
the configuration that has been obtained {\it after} passage of the worm
$w$. The antiworm thus completely undoes the action of the worm $w$, leading back to
the configuration prior to the insertion of the worm $w$. The
antiworm is inserted at the same place as $w$, and its initial head operator is exactly the inverse 
of the last worm head operator, so that its insertion probability is
$P_{\rm insert}(T_N^\dagger,\widetilde{T}_N(s'_N))$. The probability
to create the antiworm is thus

\begin{eqnarray}
P^{\bar w}&=&P_{\rm init}\cdot P_{\rm
  insert}(T_N^\dagger,\widetilde{T}_N(s'_N)) \nonumber \\
& &\times \prod_{i=1}^{N}P_{b_i}({{\bf \bar \Sigma}}_i, T_i^\dagger \rightarrow T_{i-1}^\dagger, 
l'_i\rightarrow l_i).
\end{eqnarray}
The ratio of the two probabilities is
\begin{eqnarray}
P^w/P^{\bar w}&=&\frac{P_{\rm insert}(T_0,s_1)}{P_{\rm insert}(T_N^\dagger,
\widetilde{T}_N(s'_N))} \nonumber \\
& &\times \prod_{i=1}^{N}\frac{P_{b_i}({{\bf \Sigma}}_i, T_{i-1}\rightarrow T_i, l_i\rightarrow l'_i)}
{P_{b_i}({{\bf \bar \Sigma}}_i, T_i^\dagger \rightarrow T_{i-1}^\dagger, l'_i\rightarrow l_i)}.
\end{eqnarray}
Using Eq.~(\ref{my.dl.eq}), we obtain
\begin{eqnarray}
P^w/P^{\bar w}&=&\frac{P_{\rm insert}(T_0,s_1)}{P_{\rm insert}(T_N^\dagger,\widetilde{T}_N(s'_N))} 
\nonumber \\
& &\times \prod_{i=1}^{N}\frac{f(T_{i-1},s_i) W(b_i,{{\bf \Sigma}}_i)}{f(T_i^\dagger,\widetilde{T}_i(s'_i)) 
W(b_i,{{\bf \bar \Sigma}}_i)}.\label{eq:intermediate}
\end{eqnarray}
Since $s'_i=s_{i+1}$ we obtain, using Eq.~(\ref{eq:f.hermit}),
$$
f(T^\dagger_i,~\widetilde{T}_i(s'_i))=f(T_i,s_{i+1})
$$
for all $i<N$. 
For a ``normal'' loop, we furthermore have $T_N=T_0$ and $s'_N=s_1$, so that
$$ f(T_N^\dagger,\widetilde{T}_N(s'_N))=f(T_0,s_1),$$
again using Eq.~(\ref{eq:f.hermit}). In case of  a ``bounce'' loop, where $T_N=T_0^\dagger$ and
$s'_N=\widetilde{T}_0(s_1)$, we obtain the same relation, since
$f(T_0,\widetilde{T^\dagger_0}(\widetilde{T}_0(s_1))=f(T_0,s_1).$

The factors of $f(T,s)$ thus exactly cancel each other in the numerator
and denominator of Eq.~\ref{eq:intermediate}, and we obtain
\begin{eqnarray}
P^w/P^{\bar w}&=&\frac{P_{\rm insert}(T_0,s_1)}{P_{\rm insert}(T_N^\dagger,\widetilde{T}_N(s'_N))} 
\nonumber \\
& &\times \prod_{i=1}^{N}\frac{W(b_i,{{\bf \Sigma}}_i)}{W(b_i,{{\bf \bar \Sigma}}_i)}.
\end{eqnarray}
Detailed balance is thus fulfilled, provided
\begin{equation}
\label{eq:above.cond}
P_{\rm insert}(T_0,s_1) = 
P_{\rm insert}(T_N^\dagger,\widetilde{T}_N(s'_N)).
\end{equation}

For a ``bounce'' loop, where
  $T_N^\dagger=T_0$ and $\widetilde{T}_N(s'_N)=s_1$,  this condition is always fulfilled.
In case of a ``normal'' loop, where $T_N=T_0$ and $s'_N=s_0$, we need for
  Eq. (\ref{eq:above.cond}) to hold, that
\begin{equation}
\label{eq:insert.condition}
P_{\rm  insert}(T_0^\dagger,\widetilde{T}_0(s_1))=P_{\rm insert}(T_0,s_1).
\end{equation}
In other words, the probability to insert (at the same place) an antiworm that will undo exactly what a worm just did must be equal to the probability used to insert this original worm. If this condition is fulfilled for all transformations $T_0$ and all possible
states $s_1$, we obtain a detailed balanced operator-loop update.

\subsection{Operators, insertion probabilities, worm weights, and Green's functions}
\label{sec:trans}
Let us be more specific now, and discuss the kind of operators $A_0$ that can
be used as operator insertions, and which corresponding insertion
probabilities fulfill Eq. (\ref{eq:insert.condition}). We focus on two cases: quantum spin-$S$  and softcore bosonic systems.

For a quantum spin-$S$ system, the local state at a given site is given by the projection of the
spin value at that 
site, e.g. onto the $z$ axis. We denote this projection by $m$ which can take $2S+1$ 
values in the range $-S,-S+1,\ldots,S-1, S$. 

For bosonic systems, the local state is given by the number $n$ of bosons 
at the site. If we truncate the Hilbert space by restricting the number of bosons per site 
to a maximum value $N_{\rm max}$, $n$ can take integer values in the range $0,\ldots,N_{\rm max}$.

What are the possible operators $A_0$ to be used in the operator pair insertion
for these models? 
In many cases, a good choice is to construct so called $\pm 1$ worms: A
$+1$ ($-1$) worm head acting on state $s$ changes it to $s+1$ ($s-1$). 
The operators $A_0$ associated with the
worm ends are then simply the creation (annihilation) operators 
$a^\dagger$ ($a$) for bosons, and the ladder operators $S^+$ ($S^-$)
for spins, respectively. 

\subsubsection{Insertion probabilities}
\label{subsec:insertionprobabilities}
For $\pm 1$ worms, Eq. (\ref{eq:insert.condition}) becomes
$P_{\rm insert}(a,n+1)=P_{\rm insert}(a^{\dagger},n)$ in the case of bosonic models and
$P_{\rm insert}(S^-,m+1)=P_{\rm insert}(S^+,m)$ for spin models.

For a spin-$S$ model, we cannot insert a $+1$ ($-1$) worm onto a
given initial state with $m=S$ ($m=-S$). Since we always want to create a worm in
all other cases, we propose the following insertion probabilities:
\begin{equation}
P_{\rm insert}(S^{\pm},m)=\frac{1-\delta_{m,\pm S}}{2}.
\end{equation}
If $S=1/2$, we use $P_{\rm insert}(S^{\pm},m)=\delta_{m,\mp 1/2}$ instead,
so always inserting a worm.

For a softcore bosonic model limiting the maximum number of bosons per site to
$N_{\rm max}$, we equivalently use:

\begin{eqnarray}
P_{\rm insert}(a^{\dagger},n) & = & \frac{1-\delta_{n,N_{\rm max}}}{2} \nonumber \\
P_{\rm insert}(a,n) & = & \frac{1-\delta_{n,0}}{2}.
\end{eqnarray}

For hardcore bosons ($N_{\rm max}=1$), we instead use $P_{\rm insert}
(a^{\dagger},n)=\delta_{n,0}$ and $P_{\rm insert}
(a,n)=\delta_{n,N_{\rm max}}$, thus always inserting a worm.

These insertion probabilities differ from the weight
assigned to the operators in the extended configuration space, namely the matrix
elements of these operators (see subsection~\ref{subsec:wormweight}). As
suggested in Ref.~\cite{sylju1}, it is possible to set $P_{\rm
  insert}(T,s)$ proportional to $\langle \widetilde{T}(s) | T | s \rangle$,
similar to the worm algorithm~\cite{worm}. Indeed, this choice satisfies
Eq.~(\ref{eq:insert.condition}). 

\subsubsection{Worm weights}
\label{subsec:wormweight}

In the extended configuration space, where the worm head is associated with an
operator acting on the 
local state in the world-line configuration, 
the worm weights are equal
to the matrix elements, $f(T,s)\equiv\langle \widetilde{T}(s) | T | s
\rangle$.

To be more specific, consider employing $\pm 1$ worms for a spin model. 
Then $T$ can be $S^+$ or $S^-$, so we obtain
\begin{eqnarray}
f(T,s)=f(S^{\pm},m)& = & \langle m \pm 1 | S^{\pm} | m \rangle \nonumber \\
& = & \sqrt{S(S+1)-m(m\pm 1)}. 
\end{eqnarray}

For a bosonic model with $\pm 1$ worms, $T$ is either $a$ or $a^\dagger$ and thus
\begin{equation}
f(T,s)=\left\{
\begin{array}{lllll}
f(a^{\dagger},n)& = & \langle n \pm 1 | a^\dagger | n \rangle & = &
\sqrt{(n+1)} \\
f(a,n) & = & \langle n \pm 1 | a | n \rangle & = & \sqrt{n}
\end{array}
\right.
\nonumber
\end{equation}

We note, that the operators used here (corresponding to $\pm 1$ worms) are not
unique, as we can for example also employ $\pm 2$ or $\pm 3$ worms. 

It is also possible that other choices of weights such that $f(T,s)$ is not equal to $\langle \widetilde{T}(s) | T | s
\rangle$ might lead to more efficient algorithms. Indeed, in the proof of detailed balance, the only requirement on $f(T,s)$ is Eq.~(\ref{eq:f.hermit}). However, the above choices naturally appear within the extended configuration space, 
and lead to algorithms with less bounces, as will be shown below.

\subsubsection{Stopping probability}
\label{sec:stop}

We propose to always close a worm when the worm head
returns to the insertion point.
It is possible, as noted in references~\cite{harada,smakov}, to not necessary
do so, but to offer the worm the possibility to 
continue depending on the value of the final state. As a consequence, the
worm insertion probabilities need to be changed accordingly, in order to retain detailed balance. 
It is not a priori clear which approach results in a more efficient algorithm. Only precise studies of autocorrelation times could answer this question for each specific model and set of
parameters, which is however well beyond the scope of this work. Instead we
present an intuitive argument, why we expect closing worms immediately to be more efficient:

The goal of using worm updates is the generation of large non-local changes in each
MC configuration, in order to decorrelate two consecutive measurements. 
A precise quantification of this decorrelation effect in terms of CPU time 
must take into account the worm size. 
Making a long worm and thus obtaining large decorrelation 
effects should grossly be equivalent to making two short worms with only half the decorrelation. 
However, if after the first encounter of the initial point the worm has
already resulted in large enough decorrelation, it becomes less meaningful to
continue this worm,  as we can already perform an independent measurement
instead of  
spending more CPU time for the construction of a longer worm.

\subsubsection{Measuring Green's functions}
\label{sec:green}

With the above choices, the measurement of Green's functions during the worm 
construction needs to be slightly modified in order to account for the presence of 
the explicit worm weights in the worm's propagation.
For a detailed account on how the  Green's functions  measurements are performed using heat 
bath and standard directed loops with the insertion
and stopping probabilities of Sec.~\ref{subsec:insertionprobabilities}, and \ref{sec:stop} respectively,
we refer to Ref.~\cite{dorneich}. Here, we only summarize the main point:
In the standard directed loop algorithm, the value of the Green's function measurement for a given distance (in space and imaginary time) between the worm head and the worm tail equals the product of 
the matrix element of the operator inserted at the head of the worm (which would be in our notations $\langle \widetilde{T}(s) | T | s \rangle$, where $T$ denotes again the transformation corresponding to the worm head, and $s$ the local state of the world-line configuration, onto which the worm head acts) times the matrix element inserted at its tail (in our notations $\langle \widetilde{T_0}(s_0) | T_0 | s_0 \rangle$, where $T_0$ denotes the transformation corresponding to the (static) worm tail, and $s_0$ the local state of the world-line configuration, onto which the worm tail acts).
For a detailed graphical illustration of this measurement process we refer to Ref.~\cite{dorneich}. 
The only modification to this scheme, which arises from using generalized directed loops is as follows: 
In the generalized directed loop algorithm, the propagation of the worm head fulfills detailed balance in the extended configuration space. The worm head's matrix elements are thus taken into account in the probability to obtain a given configuration (in space and imaginary time) between the worm head and the worm tail.
Therefore, in the  generalized directed loop algorithm the value of  
the Green's function measurement equals $\langle \widetilde{T_0}(s_0) | T_0 | s_0 \rangle \cdot \langle \widetilde{T_0}(s_0) | T_0 | s_0 \rangle = \langle \widetilde{T_0}(s_0) | T_0 | s_0 \rangle^ 2$. Note, that this is independent of $T$ and $s$, and involves only the value of the matrix element $ \langle \widetilde{T_0}(s_0) | T_0 | s_0 \rangle^ 2$ from the static worm tail. The 
 Green's function measurement in the  generalized directed loop algorithm thus  requires significantly less evaluations of matrix elements, or accesses to their look-up table.

If
one would furthermore choose $P_{\rm insert}(T,s)$ proportional to $\langle \widetilde{T_0}(s_0) | T_0 | s_0 \rangle$,
similar to the worm algorithm~\cite{worm} discussed in Sec.~\ref{subsec:insertionprobabilities}, the value of each 
 Green's function measurement would be equal to $1$, as for the worm algorithm~\cite{worm}. In fact, this way the matrix elements of both the worm head and tail would be accounted for explicitly during the construction of the worm and its propagation. 

 \section{Numerical strategy}
\label{sec:num}

In the preceding sections we derived generalized conditions on the
scattering probabilities
$P_{b_i}({{\bf \Sigma}}_i, T_{i-1}\rightarrow T_i, l_i\rightarrow l'_i)$,
which describe the motion of the worm head at each vertex during the worm
construction. We 
now look for solutions of Eq. (\ref{my.dl.eq}), for which the bounce probability for each possible
vertex configuration is as small as possible. We expect this to lead to
an optimal algorithm in terms of autocorrelation times.
Here, we explain how to numerically solve Eq. (\ref{my.dl.eq}) for such probabilities.
Note, that the numerical procedure outline below also 
applies to the standard directed loop approach, by simply  setting $f(T,s)$ equal to $1$.

For a given vertex configuration we can construct from the scattering
probabilities $P_{b_i}({{\bf \Sigma}}_i, T_{i-1}\rightarrow T_i,
l_i\rightarrow l'_i)$ a  $4 \times 4$ ``scattering matrix'' $P$, whose
elements are 
$$
P_{kl}=P_{b_i}({{\bf \Sigma}}_i, T_{i-1}\rightarrow T_i,
l\rightarrow k),
$$
so that the element $P_{kl}$ corresponds to the probability of exiting from
leg $k$, given that the worm head entered the vertex on leg $l$.

There are various constraints on the matrix $P$. In particular,
Eq.~(\ref{my.dl.eq}) constraints the elements of $P$ according to detailed balance.
Furthermore, in order to be
interpreted as probabilities, all the matrix elements of $P$ must be contained
within $[0,1]$, that is to say 
\begin{equation}
\label{eq.P.constraint2}
0\leq P_{kl} \leq 1 \; \; \;  \forall \; k,l.
\end{equation}
Since the worm always leaves a vertex, we must have
\begin{equation}
\label{eq.P.constraint3}
\sum_k P_{kl} = 1 \; \; \;  \forall \; l,
\end{equation}
i.e. each column of $P$ must be normalized to $1$.

It is possible to add additional symmetry constraints on $P$. 
While these are not necessary conditions, they
might increase the numerical accuracy in looking for the matrix $P$. 
Given an entrance leg $l$, let us call two legs $k$ and $h$
{\it equivalent}, $k\sim h$, if the product $f\cdot W$ on the right
hand side of Eq.~(\ref{my.dl.eq}) gives the same value, independent of
choosing $k$ or $h$ as the exit leg. If two equivalent legs $k$ and $h$ both
differ from the entrance leg $l$, they must be  chosen as the exit leg
with equal probability, i.e.
\begin{equation}
\label{eq.P.constraint4}
P_{kl} = P_{hl} \; \; \; \forall \; l\neq k,h,\; k\sim h.
\end{equation}
A similar condition can be derived for equivalent entrance legs by
consideration of the reversed process.

After characterizing the constraints on the scattering matrix $P$, we can now
formulate our optimization criterion in terms of $P$.
Our goal is to construct an optimal directed-loop update, and as argued
before~\cite{sylju1,sylju2,harada}  
we aim to minimize the number of bounce processes, i.e. the bounce probabilities. In our $P$-matrix 
language, this means that we need to minimize all {\it diagonal} matrix
elements. In order not to introduce any additional 
bias among the different bounce probabilities, we require for the actual implementation 
to minimize the trace of the matrix $P$, thereby  treating all bounce probabilities equally,
\begin{equation}
\label{eq.P.min}
{\rm minimize} \; \; \; \sum_l P_{ll}.
\end{equation}

In previous studies~\cite{sylju1,sylju2,harada,smakov,zyubin}, sets of
probabilities satisfying all these conditions were obtained analytically for
specific models. From the constraints~(\ref{my.dl.eq},\ref{eq.P.constraint2}, \ref{eq.P.constraint3},\ref{eq.P.constraint4}) 
and the optimization goal~(\ref{eq.P.min}), we see that we arrive in front
of a 
{\it linear programming} problem for each scattering matrix $P$~\cite{alet03b}. This can be be solved numerically using
  standard linear programming routines~\cite{linear}. In most cases, we found
  that at most one diagonal matrix element was non zero. We also note here
  that the linear programming routines picks one of the many possibly
  equally optimal (with respect to condition~(\ref{eq.P.min})) solutions 
  depending on its initial search point. This issue will be further discussed later.

This direct way of looking for the optimal (in terms of bounce minimization) solutions of the directed loop equations 
is not specific to any model and needs no preceding analytical calculation.
It allows for a rather generic implementation of the SSE algorithm, where
after implementation of the Hamiltonian, a standard  minimization
routine~\cite{linear} can be employed  in
order to obtain the scattering matrices prior to starting the actual simulation. 

\section{Algorithmic phase diagrams}
\label{sec:phased}

In this section, we apply the preceding method to the
simulation of quantum systems which have  
been extensively studied previously using the SSE QMC method. 

To ensure that all diagonal matrix elements of the bond Hamiltonians are
positive, we add a constant $C=C_0+\epsilon$ per bond to the original
Hamiltonian where $C_0$ is the minimal
value for which all diagonal matrix elements are positive, and $\epsilon \geq
0$. We will see that usually a finite value of $\epsilon$ is required in order to allow for regions in parameter space which are completely bounce-free. In general, we find that increasing $\epsilon$ results in lower bounce 
probabilities. However, as the size of the operator string grows with 
$\epsilon$, this leads to increasing simulation times: there is clearly a tradeoff between more bounces but less CPU time (small $\epsilon$) and less bounces but more CPU time (large $\epsilon$). 
We expect that there is no general rule how to a priori choose the  
value of $\epsilon$ in order to obtain the smallest autocorrelation times.

\subsection{Heisenberg model}

First we consider the  easy-axis spin-$S$ Heisenberg model in an
external magnetic field $h$,
\begin{equation}
H=J \sum_{\langle i,j \rangle} \frac{1}{2}(S^+_iS^-_j + S^-_iS^+_j) + \Delta
S^z_i S^z_j - h \sum_i S^z_i,
\label{eq:HSM}
\end{equation}
where $\Delta$ denotes the easy-axis anisotropy, and the first sum extends
over all nearest neighbors on the $d$-dimensional hypercubic lattice.

\begin{figure}
\includegraphics[width=7cm]{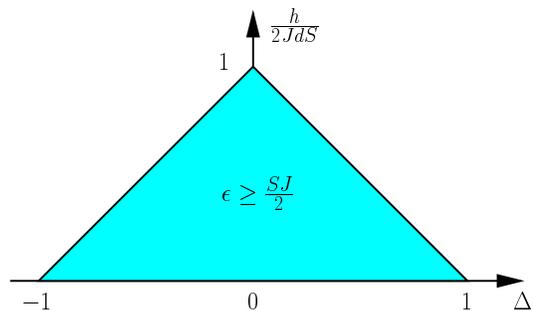}
\caption{(Color online) Algorithmic phase diagram for an easy-axis anisotropic spin-$S$ Heisenberg model
  in a magnetic field $h$ field, on a $d$-dimensional cubic lattice with nearest neighbor exchange $J$. The
  easy-axis anisotropy is denoted $\Delta$. The shaded region indicates those parameters,
  for which a bounce-free solution of
  the generalized directed loop equations can be found.}
\label{fig:AlgoS}
\end{figure}

Numerically scanning the parameter space ($\Delta,h\geq0$), we search for regions
where our optimization procedure finds bounce-free solutions (i.e. $\sum_l P_{ll}=0$ for all allowed vertices).
From this procedure we obtain the algorithmic phase diagram displayed in Fig.~\ref{fig:AlgoS}.

We find a finite region of the parameter space ($|\Delta |+h/(2dS)\leq J$)
which corresponds to bounce-free solutions of the generalized directed-loop
equations. Within this region typically one needs $\epsilon \geq
SJ/2$. However, for $\Delta = \pm J$, we  find  bounce-free solutions also for
$\epsilon=0$.
Outside the bounce-free region at least one of the scattering matrices does not allow for a
traceless solution. 

For $S=1/2$, Sylju\aa sen and Sandvik analytically found the same bounce-free
region~\cite{sylju1}. By monitoring the parameter dependence of
the finite bounce probabilities, we verified that our numerical approach
indeed yields their analytical solution.

Sylju\aa sen recently extended the directed loop framework proposed
in~\cite{sylju1} to spin-$S$ models~\cite{sylju2}. 
Within our framework, his ansatz corresponds to setting $f(T,s)=1$.
He finds no region in
parameter space where the directed loop equations allow for bounce-free solutions
for any $S>1$. 
We have verified this by setting $f(T,s)=1$ and find that for $S>1$, there is
indeed no bounce free solution, using Sylju\aa sen's choice.
For $S=1/2$ and $S=1$, Sylju\aa sen recovers the phase diagram
shown in Fig.~\ref{fig:AlgoS}. 
This reflects the fact that for $S=1/2$ and $S=1$ all non-zero matrix elements of
the $S^\pm$ operators are equal and thus the factors $f(T,s)$
cancel out of the generalized directed loop equations, making our
generalization equivalent to the standard approach.
For $S>1$ the generalized directed loop equations, including the worm weights
as extra degrees of freedom, however allow for more bounce free solutions.

The algorithmic phase diagram shown in Fig.~\ref{fig:AlgoS} was also found to hold
for the coarse-grained loop algorithm~\cite{harada}. 
This suggests that the numerically determined
scattering probabilities are similar to 
those of the coarse graining approach. This equivalence is also pointed out
more clearly in Ref.~\cite{kawa.lanl}.

\subsection{Softcore Bosonic Hubbard model}

Here we present algorithmic phase diagrams for the bosonic Hubbard
model, with Hamiltonian
\begin{equation}
H=-t\sum_{\langle i,j \rangle} a^\dagger_i a_j + a_i a^\dagger_j + U/2\sum_i
n_i(n_i-1)-\mu \sum_i n_i,
\end{equation}
where the $a_i^{\dagger}$ ($a_i$) denote boson creation (destruction) operators on
sites $i$, $n_i=a^{\dagger}_ia_i$ is the local density , $t$ is hopping amplitude, $U$ the on-site interaction, and $\mu$
the chemical potential.

We need to restrict the simulation to a maximum number
  $N_{\rm max}>1$ of bosons per lattice site, in order to obtain positive
  diagonal bond Hamiltonian matrix elements. For the hardcore bosonic case
  $N_{\rm max}=1$, we refer to the preceding section, since the hardcore
  bosonic Hubbard model exactly maps onto
  a spin-$1/2$ antiferromagnetic Heisenberg model. 

\begin{figure}
\includegraphics[width=7cm]{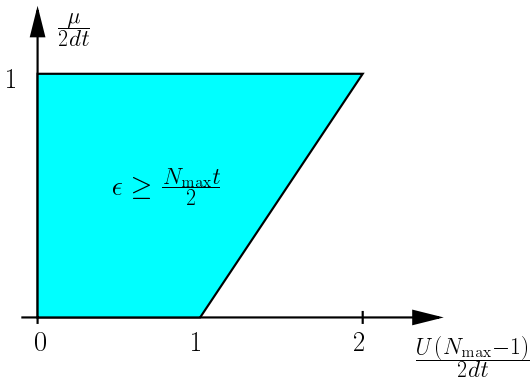}
\caption{(Color online) Algorithmic phase diagram for the bosonic Hubbard model on a
  $d$-dimensional cubic lattice with nearest neighbor hopping $t$, onsite interaction strength $U$, and a chemical potential $\mu$. $N_{\rm max}$ denotes the 
cutoff in the local occupation number. The
  shaded region indicates the regime of bounce-free solutions of the generalized
  directed loop equations.}
\label{fig:AlgoB}
\end{figure}

Using our numerical optimization technique we arrive at the algorithmic phase diagram
shown in Fig.~\ref{fig:AlgoB}. There is a finite region of bounce-free
solutions to the directed loop equations. However, this region shrinks upon increasing
$N_{\rm max}$, and we need to allow $\epsilon \geq N_{\rm
  max}t/2$ in order to recover the complete bounce free region.

Sylju\aa sen studied the directed loop equations for bosonic
models and did not obtain bounce-free regions for any  $N_{\rm max}>1$~\cite{sylju2}.
The same result was obtained in Ref.~\cite{zyubin}. 
Again this indicates the importance of allowing the additional weight factors within our approach.

Smakov {\it et al.}~\cite{smakov} presented a coarse-grained loop algorithm for the 
simulation of softcore bosons. They present results for
free bosons, for which no constraint on the occupation number is necessary
within the SSE approach. Since their method proceeds directly in the $N_{\rm max} \rightarrow \infty$ limit, 
we expect that using their algorithm there will remain no bounce-free regions for finite
on-site interaction.

For a softcore bosonic model without a cutoff on the
maximum value of bosons per site, it is also possible to perform simulations by imposing an
initial cutoff $N_{\rm max}$, which is then adjusted during the course of the thermalization process.
With the numerical procedure at hand, it is easy to recalculate the scattering
matrices $P$ when needed, namely when the current cutoff becomes too small, and needs to be increased.

\section{Autocorrelation results}
\label{sec:auto}

The results presented in the  previous section suggest that
the generalized directed loop equations lead to efficient update
schemes. In particular, in many cases we could greatly extend 
bounce-free regions in parameter space using the generalized directed loop method.

It is generally expected that reducing bounce processes leads to more
efficient algorithms.
In this section, we therefore compare the efficiency of an arbitrarily picked solution to the generalized
directed loop algorithm to earlier approaches:
the original heat bath choice for the scattering probabilities by
Sandvik~\cite{loopoperator}, and the
directed loop approach by Sylju\aa sen and
Sandvik~\cite{sylju1,sylju2}. This
comparison is performed using 
the example of the magnetization process of quantum spin
chains. 

We define each MC step to consist of a full diagonal update,
 followed by a fixed number $N_w$ of worms updates, where $N_w$ is chosen such that on average twice
 the number of vertices in the operator string are hit by those worms. We perform a
 measurement after each such Monte Carlo step, and determine integrated autocorrelation
 times using standard methods~\cite{evertz}.

In case the effort for a single MC step was the same for each of the three
algorithms, the integrated autocorrelation time would
establish a valid comparison between these algorithms in terms of CPU time. 
Suppose, however, that a MC step of Alg.~A took twice the CPU time than a
MC step using Alg.~B. In that case even with a 50\% reduction of the
autocorrelation time upon using Alg.~A, both would be equally efficient, since
in order to obtain a given number of independent configurations, the same CPU
time would be needed.
In the following, we therefore present a measure of autocorrelations, which
takes the effort of each update scheme into account in a machine independent way.

For this purpose, we  define the worm size $w$ as the total number of vertices that
  have been visited by the worm, including those visited during bounce
  processes~\cite{harada,alet03b}. 
The number $N_w$, calculated self-consistently during thermalization, is
  then defined such that ${N_w} \langle w \rangle \sim 2 \langle
  n \rangle$, where $\langle n \rangle$ is the average number of non-identity
  operators in the operator string ($\langle \ldots \rangle$ denotes MC
  averages). 
In counting $N_w$ we  include worms that are immediately stopped.
The number $N_w$ can fluctuate from one simulation to another, and more importantly
depend on the underlying algorithm: indeed, the worms constructed using different algorithms are
not expected to be of the same size.
In order to account for this difference in effort, we 
multiply the integrated autocorrelation times by a factor
 $N_w \langle w \rangle/{\langle n\rangle}$, which is close to $2$ by
  definition, 
but which might differ, depending on the underlying algorithm.
 
The results presented below were obtained by the following
procedure: for each of the three algorithms  we run
simulations containing  $10^6$ MC steps and
calculate integrated autocorrelation times $\tau_0$ for various
observables~\cite{evertz}. 
From these we obtain effort-corrected autocorrelation times $\tau=
\tau_0 \cdot N_w \langle w \rangle/{\langle n\rangle}$, leading to a
machine-independent measure of efficiency.
We applied the above procedure to the autocorrelations of the uniform magnetization and energy
of antiferromagnetic spin chains in finite magnetic fields, and present
results for the spin-$3/2$ $XY$ and the spin-$2$ Heisenberg case.

\subsection{Spin-$3/2$ $XY$ chain}
\label{sec:spin32}

\begin{figure}
\includegraphics[width=8cm]{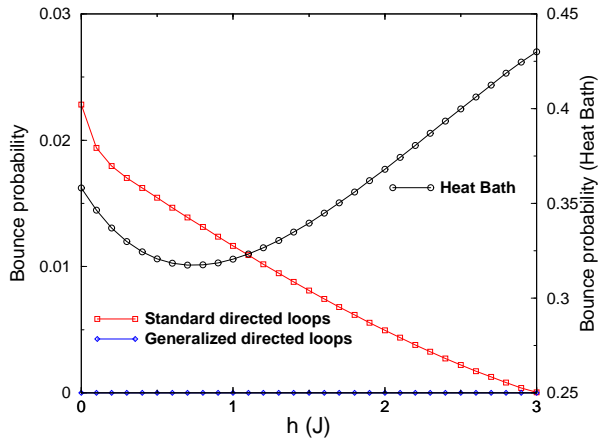}
\caption{(Color online) Bounce probabilities using the heat bath algorithm, standard
  directed loops, and generalized directed loops for a $L=64$ sites spin-$3/2$ $XY$
  chain in a magnetic field $h$ at $\beta=64/J$. A different scale is used for the heat bath algorithm.}
\label{fig:Bounce.Proba.XY}
\end{figure}

We simulated the spin-$3/2$ $XY$ model (Eq. (\ref{eq:HSM}) with $\Delta=0$) on a $L=64$ sites chain at an 
inverse temperature
$\beta=64/J$, for fields from zero up to saturation, $h=3J$. 
For the simulations presented here, we chose $\epsilon=SJ/2-h/4$ which is
found to be the minimal value to have a bounce free algorithm for a $XY$ chain in a field.
The magnetic field dependence of the bounce probability is shown
for all three algorithms in Fig.~\ref{fig:Bounce.Proba.XY}.
The bounce probability is rather large ($30-45 \%$ for all fields) for the heat bath algorithm
and significantly reduced (to less than 2\%) using the standard directed loop equations, while it vanishes 
all the way up to the saturation field using generalized directed loops.

\begin{figure}
\includegraphics[width=7cm]{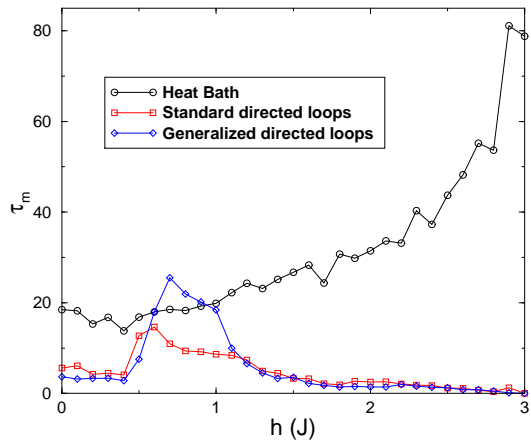}
\caption{(Color online) Autocorrelation times for the uniform magnetization, measured using
  heat bath, standard directed loops, and generalized directed loops for a $L=64$ sites 
spin-$3/2$ $XY$
  chain in a magnetic field $h$ at $\beta=64/J$.}
\label{fig:TauM.XY}
\end{figure}

\begin{figure}
\includegraphics[width=7cm]{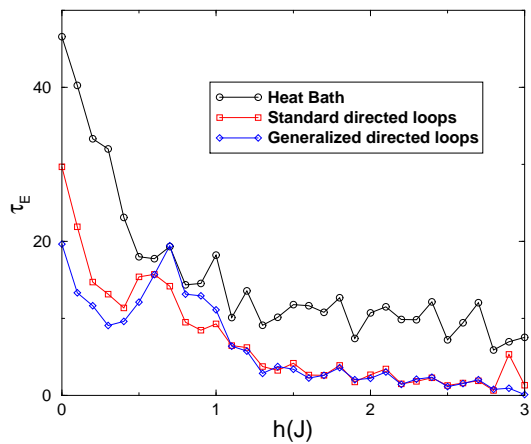}
\caption{(Color online) Autocorrelation times for the energy, measured using heat bath, standard directed loops, and 
generalized directed loops for a $L=64$ sites spin-$3/2$ $XY$
  chain in a magnetic field $h$ at $\beta=64/J$.}
\label{fig:TauE.XY}
\end{figure}

The rescaled autocorrelation times of the magnetization ($\tau_m$) and 
energy ($\tau_E$) are shown as functions of the magnetic field strength in
Figs.~\ref{fig:TauM.XY} and~\ref{fig:TauE.XY}, respectively. 
Using the heat bath algorithm, $\tau_m$ increases upon increasing $h$, while
$\tau_E$ decreases. The uniform magnetization of the MC configuration is
updated only during the operator-loop updates, while the energy is not changed
during this update step \cite{loopoperator}. Therefore autocorrelations in the energy measurements are
less sensitive to the efficiency of the operator loop update, and mainly decrease
with field strength, due to increasing operator string lengths.
In both the low and the high field region, the improvements of standard and generalized directed loops upon using the heat bath algorithm are clearly seen for both the energy and magnetization in Figs.~\ref{fig:TauM.XY} and~\ref{fig:TauE.XY}. Within our scheme, we find small but not significant improvements over the standard directed loops, and for $h\sim J$, the bounce-free solution even results in slightly larger autocorrelation times than the heat bath method.

This clearly indicates that one must include further strategies, besides the bounce minimization in order to obtain a better algorithm, as will be discussed in Sec.~\ref{sec:strat}.

\subsection{Spin-$2$ Heisenberg chain}

\begin{figure}
\includegraphics[width=8cm]{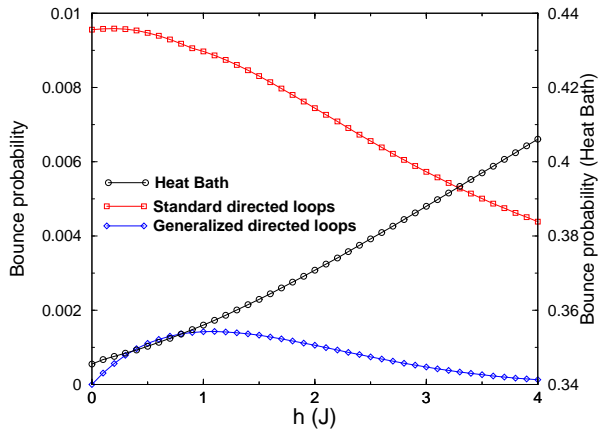}
\caption{(Color online) Bounce probabilities using the heat bath algorithm, standard
  directed loops, and generalized directed loops for a $L=64$ sites
  spin-$2$ antiferromagnetic Heisenberg
  chain in a magnetic field $h$ at $\beta=64/J$. 
  A different scale is used for the heat bath algorithm.}
\label{fig:Bounce.Proba.AF}
\end{figure}

Next, we consider the isotropic ($\Delta=1$) antiferromagnetic spin-$2$ Heisenberg model in a magnetic
field. We simulated a chain with $L=64$ sites at $\beta J = 64$ and for 
fields ranging from  zero up to saturation at $h=4J$, and using $\epsilon=SJ/2$. In
Fig.~\ref{fig:Bounce.Proba.AF},  the resulting bounce probabilities are shown as 
functions of magnetic field strength for the three algorithms.
Similar to the previous case, the bounce probabilities are rather high using heat bath,
  $\sim 34-42\% $), whereas they are significantly reduced using 
the directed loop algorithms (less than $1\%$ in both cases). 
Even though the bounce probabilities are finite at $h>0$ for the generalized directed
loop algorithm, they are smaller than for the
standard directed loop algorithm. Furthermore,  
in the limit of zero field, using generalized directed loops 
leads to a vanishing bounce probability.

\begin{figure}
\includegraphics[width=7cm]{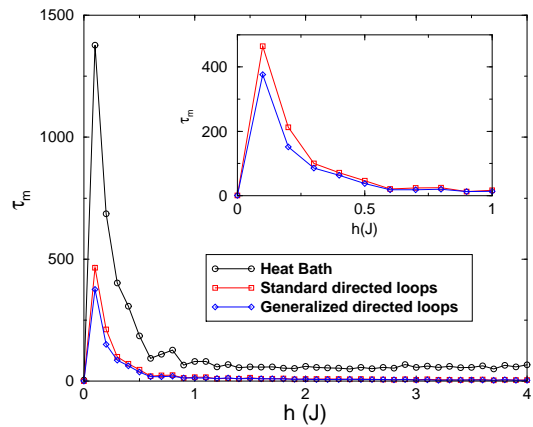}
\caption{(Color online) Autocorrelation times for the uniform magnetization, measured using the
  heat bath algorithm, standard directed loops, and generalized directed loops for a $L=64$ sites 
spin-$2$ 
antiferromagnetic Heisenberg
  chain in a magnetic field $h$ at $\beta=64/J$. The inset shows on a larger
  scale the autocorrelation times using directed loops at small values of the field.}
\label{fig:TauM.AF}
\end{figure}

\begin{figure}
\includegraphics[width=7cm]{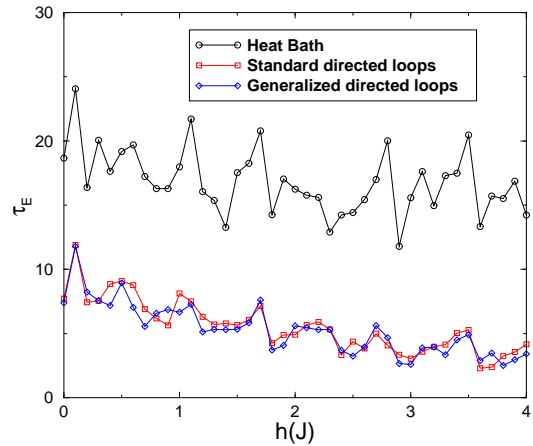}
\caption{(Color online) Autocorrelation times for the energy, measured using heat
  bath, standard directed loops, and generalized directed loops
for a $L=64$ sites spin-$2$ antiferromagnetic Heisenberg
  chain in a magnetic field $h$ at $\beta=64/J$.}
\label{fig:TauE.AF}
\end{figure}

In Fig.~\ref{fig:TauM.AF} and Fig.~\ref{fig:TauE.AF}, we present results
for rescaled autocorrelation times of the magnetization ($\tau_m$) and energy
($\tau_E$).
The dependence of $\tau_m$ on the magnetic
field has a similar tendency for all three algorithms: starting from a small value at zero
field, $\tau_M$ is sharply peaked at $h \sim 0.1 J$, and
decreases rapidly upon further increasing the field strength, reaching an almost constant value. 
This sharp peak around $h \sim 0.1 J$ probably corresponds to the closure of
the Haldane gap (estimated as $\Delta_H=0.08917(4) J$ for the spin 2
chain~\cite{todo}) by
the magnetic field.
We observe that $\tau_M$ is larger by
nearly a factor of 3 using heat bath rather than directed loops.
This is expected given the larger bounce
probabilities in Fig.~\ref{fig:Bounce.Proba.AF}.
We find that independent of the magnetic field strength,
$\tau_m$ is less for the generalized directed loop algorithm than for the standard one.

Concerning the autocorrelation times $\tau_E$ shown in Fig.~\ref{fig:TauE.AF}, we reach similar
conclusions as for the spin-$3/2$ $XY$ case: the autocorrelation
times of the energy are reduced by a factor around 2 from those using the heat bath algorithm.

\section{Optimizing directed loop algorithms}

\label{sec:strat}

The results in the previous section clearly indicate that minimizing bounces alone is not sufficient to obtain an efficient algorithm, since the bounce-free (or bounce-minimized) solution is not unique~\cite{sylju2,zyubin}. The numerical lineal programming solver employed picks a particular solution, which might not be
the optimal one in terms of autocorrelations.
In this section, we present supplementary strategies
aiming at locating more efficient solutions. 
We note that these strategies are not  specific to the generalized directed loop scheme presented in the previous sections, but can also be used to optimize the standard directed loop approach~\cite{sylju1,sylju2}.

\subsection{Supplementary strategies}

\begin{figure}
\includegraphics[width=7cm]{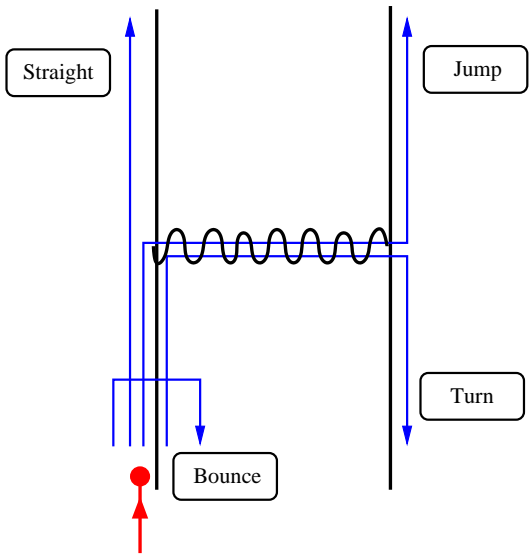}
\caption{(Color online) The different paths a worm can take across a vertex:
  "Bounce", "Jump", "Straight" or "Turn".}
\label{fig:paths}
\end{figure}

Apart from the "bounce" path, where the worm backtracks, there are three other
paths that a worm can take across a vertex. We denote
these other paths as "jump", "straight" and "turn"~\cite{dorneich}. 
See Fig.~\ref{fig:paths} for an illustration of these definitions.

Once the bounces have been minimized or even eliminated, one might consider the effects of
the remaining three paths of the worm-scattering process on the autocorrelation times.
A practical means of doing so is as follows: First, we use linear programming to 
minimize the bounces
(Eq.~\ref{eq.P.min}) and to obtain for each vertex configuration the lowest value of
the bounce (denoted $b$, where $b$ can take a different value for
each possible vertex). 
In a second step, we then {\it impose} the condition
$\sum_l P_{ll} = b$ as a new constraint, in addition to
Eqs. (\ref{eq.P.constraint2}, \ref{eq.P.constraint3},\ref{eq.P.constraint4}), so that
any feasible solution will be in the optimal subspace with respect
to bounce minimization. 

We then consider new optimization goals, 
each chosen from the 
six following possibilities: we could minimize or maximize the jump, straight or
turn probabilities. The jump probabilities simply correspond to the  scattering matrix
elements $P_{14}$, $P_{41}$, $P_{23}$, $P_{32}$, the straight probabilities to
$P_{13}$, $P_{31}$, $P_{24}$, $P_{42}$ and the turn probabilities to
$P_{12}$, $P_{21}$, $P_{34}$, $P_{43}$.
For each of the six different strategies, we use linear programming 
with the additional constraint to minimize or maximize the sum
of these matrix elements for each vertex configuration. Then we 
use the resulting scattering matrices in the SSE algorithm.
Note, that due to the additional constraint, we explicitly ensure
that these algorithms will have a minimal number of bounces.
Doing so, we obtain six sets of scattering matrices, 
each corresponding to one of the above optimization goals.

\subsection{Results}

As an example of testing the efficiency of these strategies, we consider 
the $S=3/2$ $XY$ chain in the parameter regime, where  
we found the generic solution of the generalized directed loop equations in Sec.~\ref{sec:spin32}
to perform worse than the heat bath solution.
In particular, we consider a chain with $L=64$, $\beta J=64$, $\epsilon=SJ/2$, and a value of the 
magnetic field $h=0.6J$.

In Tab.~\ref{tab:auto} we present results for the autocorrelation times of the
magnetization ($\tau_M$), staggered magnetization ($\tau_{M_{s}}$) and
energy ($\tau_{E}$) from using  each of the six different strategies.

\begin{table}
\caption{\label{tab:auto}
Autocorrelation times for the uniform magnetization ($\tau_M$), the staggered magnetization ($\tau_{M_{s}}$), and
the energy ($\tau_E$) for the generalized directed loop algorithm applied to a $L=64$ sites spin-$3/2$ $XY$
chain in a magnetic field $h=0.6$ at $\beta=64/J$,
obtained using algorithms where supplementary
strategies have been used after minimization of bounces, as explained in the
text, for $\epsilon=3/4J$.
}
\begin{ruledtabular}
\begin{tabular}{lrrr}
Supplementary Strategy & $\tau_M$ & $\tau_{M_{s}}$ & $\tau_E$ \\
\hline
Maximize Jump & 2.9 & 20.4 & 6.4\\
Minimize Jump & 22.9 & 12.5 & 16.9\\
Maximize Straight & 2.9 & 6.4 & 9.4\\  
Minimize Straight & 12.4 & 22.5 & 13.2\\   
Maximize Turn & 45.7 & 22.4 & 25.2\\
Minimize Turn & 2.7 & 23.6 & 6.6\\
\end{tabular}
\end{ruledtabular}
\end{table}

The subspace of bounce-free solutions contains algorithms with autocorrelation times
varying by about an order of magnitude; this indicates that a solution taken from this subspace 
without further guidance in general will not be the optimal one.

From Tab.~\ref{tab:auto} we furthermore find, that the optimal additional strategy depends
on the observable of interest. For example, in order to minimize the
autocorrelation times of the energy, maximizing jumps is more efficient  
than maximizing the straight path, whereas 
for the staggered magnetization
the two strategies perform opposite.
This indicates, that in general it will {\it not} be possible to obtain a 
unique optimal strategy beyond the minimization of bounces.

Minimizing bounces appears reasonable from an algorithmic point of view,
in order to prevent undoing previous changes to a QMC configuration. However, 
autocorrelations are also related to the physical phases
of the model under consideration, and thus less well captured by a generic 
local prescription for the worm propagation. 
In practice, the most efficient way to proceed for a given model will be to perform
simulations for each different strategy on small systems, in order to determine the optimal
strategy for the observable of interest
before performing production runs on larger systems.

\begin{figure}
\includegraphics[width=8.9cm]{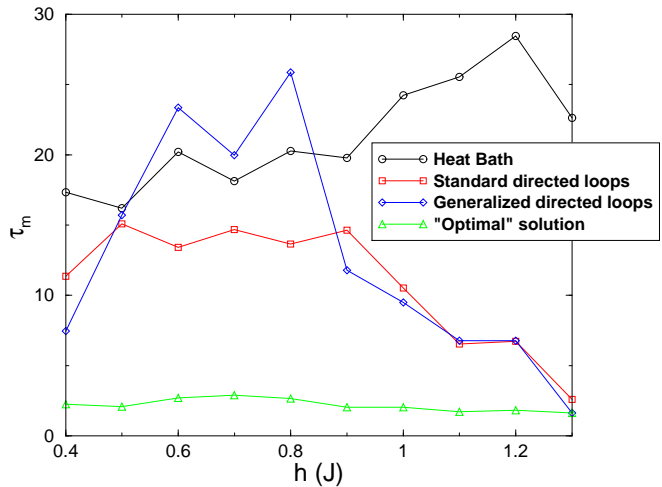}
\caption{(Color online) Autocorrelation times for the uniform magnetization, measured using
  heat bath, standard directed loops, generalized directed loops, and 
  the optimal solution (see text) for a $L=64$ sites spin-$3/2$ $XY$ chain in a
  magnetic field $h$ at $\beta=64/J$, for $\epsilon=3/4J$.}
\label{fig:TauM.optimal}
\end{figure}

In order to illustrate the reduction of autocorrelation times that can be achieved using this scheme,
we finally consider the spin-$3/2$ $XY$ chain throughout the whole region of magnetic fields,
$h=0.4J-1.3J$, where we found the unexpected increase in the autocorrelation
times (see Fig.~\ref{fig:TauM.XY}). 
The resulting minimal autocorrelation times 
for the magnetization
are shown in  Fig.~\ref{fig:TauM.optimal},
along with the
results for the  autocorrelation times using heat-bath,
standard and generalized directed loops (without additional
constraints).
Our results clearly demonstrate, that
the optimal algorithm gives rise to much better performance,
in particular  
curing the  autocorrelation time anomaly found in the
previous section. 
We find that the optimal strategy depends on the magnetic field strength: for example, we find the best
strategy to be (i) maximizing jumps for fields strengths $h=0.4J,0.5J,0.7J,1.1J$, and $1.3J$,
(ii) minimizing turns for  $h=0.6J,1.0J$, and $1.2J$, and (iii) maximizing straight moves for
$h=0.8J$ and $0.9J$. 

\section{Conclusion}
\label{sec:conc}

In this paper we presented a generalized approach to the
construction of directed loops in quantum Monte Carlo simulations. Viewing the worms ends not as artificial discontinuities, but as physical operators with corresponding weights we arrived at generalizations of the directed loop equations.
Using linear  programming techniques to solve these equations we can avoid the analytical calculations needed in previous approaches, and arrive at a generic QMC algorithm.

The generalized directed loop equations allow bounce-free solutions in larger regions
of parameter space, but measurements of autocorrelation times for several models showed that minimizing bounces is not always sufficient to obtain an efficient algorithm.

We therefore proposed a different means of further optimizing  directed loop algorithms inside the
subspace of bounce-minimal solutions. Additional strategies were presented, the use of which improves the performance
up to an  order of magnitude. However, the optimal strategy  in general depends on both the model {\it and} the observable 
of interest.
One therefore needs to perform preliminary simulations 
to find out which supplementary strategy is optimal for a given problem
before turning to long calculations, in order to account for the physical
phase realized in the specific parameter regime.

A recent paper~\cite{pollet} discussed issues similar to the ones addressed here: can one obtain strategies that improve the efficiency of QMC algorithms beyond the directed loop scheme~? In our understanding of their work, the authors of Ref.~\cite{pollet} propose to {\it always} keep a non-zero bounce probability to vertices with the largest weight. They then  provide a precise form of the scattering matrices. In Ref.~\cite{sylju2}, Sylju\aa sen also proposed to keep a non-zero bounce probability for the vertex with the largest weight in situations where he did not find bounce-free solutions. The main difference between the approach of Pollet {\it et al.}, and Sylju\aa sen thus concerns the off-diagonal elements of the scattering matrix. As shown explicitly in Sec.~\ref{sec:strat}, the off-diagonal matrix elements strongly affect the efficiency of the algorithm in a parameter- and observable-dependent way. This  indicates that there will be no simple rule for the construction of the scattering matrices, which perform optimal in all cases. Similar conclusions were reached in Ref.~\cite{pollet}. 
A full SSE code featuring the implementation of the generalized directed loop
technique described in the present paper is available as part
of the ALPS project~\cite{ALPS}.

\begin{acknowledgments}
We thank K. Harada, N. Kawashima, A. Sandvik, E. S\o rensen, O. Sylju\aa sen
and S. Todo for fruitful discussions. The simulations were done using the ALPS libraries~\cite{ALPS} and performed on the Asgard
Beowulf cluster at ETH Z\"urich. This work is supported by the Swiss National Science Foundation.

\end{acknowledgments}


\begin{references}
\bibitem{note.worm} In principle, both head and tail of the worm can move
  freely, but in this work, we restrict ourselves for simplicity to the case
  where the tail remains still.
\bibitem{notations} Formally, we should write the initial state as $|s_1\rangle$ and the 
new state as $~\widetilde{T}_0|s_1\rangle$, but for clarity reasons, we use  the simpler notations $s_1$ and $~\widetilde{T}_0(s_1)$.
\end{references}
\end{document}